%% file: main-preprint.tex
\documentclass{preprint}
\usepackage{tabularray}
\usepackage{booktabs}
\usepackage{graphicx}
\usepackage[colorlinks=true]{hyperref}
\usepackage[numbers,comma,sort&compress]{natbib} 
\usepackage[table]{xcolor}
\usepackage{tabularx}
\usepackage{xspace}
\usepackage{pifont}
\usepackage{pdflscape}

\usepackage{enumitem}
\usepackage{subcaption}
\usepackage[capitalize]{cleveref}

\newcommand{\modela}{Setting~A\xspace}
\newcommand{\modelb}{Setting~B\xspace}
\newcommand{\cmark}{\ding{51}}%
\newcommand{\xmark}{\ding{55}}%

\definecolor{stone}{HTML}{BFBA9C}
\definecolor{grey}{HTML}{F6F3EB}

\NewTblrTheme{thesis}{
    \SetTblrStyle{caption}{hang=0pt}
    \SetTblrStyle{caption-tag}{font=\bfseries}
    \SetTblrStyle{caption-sep}{font=\bfseries}
    \SetTblrStyle{capcont}{hang=0pt}
    \SetTblrStyle{capcont-tag}{font=\bfseries}
    \SetTblrStyle{capcont-sep}{font=\bfseries}
}

\title{Fracture Risk Prediction in Adults Over 50 Years Old Using DXA and EHR: Comparison of Traditional and Machine Learning Models in Two Large Cohorts}
\author[1,$\dagger$]{Jiahe Qian}
\author[2,$\dagger$]{Hao Dai}
\author[1]{Kunyu Yu}
\author[1]{Hexin Dong}
\author[2]{Xing He}
\author[3,4]{Erik A. Imel}
\author[2,4,*]{Jiang Bian}
\author[1,*]{Yifan Peng}
\author[5,*]{Yi Liu}

\affil[1]{Department of Population Health Sciences, Weill Cornell Medicine, New York, USA}
\affil[2]{Department of Biostatistics \& Health Data Science, Indiana University School of Medicine, Indiana, USA}
\affil[3]{Department of Medicine, School of Medicine, Indiana University, Indiana, USA}
\affil[4]{Center for Biomedical Informatics, Regenstrief Institute, Indiana, USA}
\affil[5]{Department of Medicine, Weill Cornell Medicine, New York, USA}

\affil[$\dagger$]{These authors contributed equally to this work.}
\affil[*]{Corresponding author(s). Email(s): \url{bianji@regenstrief.org}, \url{yip4002@med.cornell.edu}, \url{yil9015@med.cornell.edu}}

\begin{document}

\maketitle

\begin{abstract}
Accurate fracture risk prediction is important for osteoporosis management, but commonly used clinical tools may not fully use information available in electronic health records (EHRs) and dual-energy X-ray absorptiometry (DXA) reports. We developed and externally validated time-to-event fracture prediction models among adults aged 50 years or older with clinically obtained DXA reports in 2 US health care systems. The development cohort was derived from NewYork-Presbyterian/Weill Cornell Medical Center and the external validation cohort from the Indiana Network for Patient Care. Predictors included demographics, lifestyle factors, prior fracture, comorbidities, medication exposures, osteoporosis treatment history, and DXA-derived T-scores extracted from radiology reports. The outcome was time from index DXA to first incident fragility fracture identified from structured diagnosis codes. We evaluated penalized Cox regression, random survival forest, gradient-boosting survival, and XGBoost survival models using 2 prespecified predictor settings and compared discrimination with clinically reported FRAX major osteoporotic fracture probabilities. The development cohort included 11,510 adults, of whom 858 sustained incident fragility fractures; the external validation cohort included 1,932 adults, of whom 180 sustained fractures. In internal validation, the expanded Cox model achieved a mean Harrell C-index of 0.779 (95\% CI, 0.770-0.788), compared with 0.653 (95\% CI, 0.646-0.660) for FRAX. In external validation, the corresponding Cox model achieved a Harrell C-index of 0.714 (95\% CI, 0.662-0.765), compared with 0.590 (95\% CI, 0.544-0.637) for FRAX; gradient-boosting survival had the highest external discrimination (0.725; 95\% CI, 0.675-0.776). EHR- and DXA-enhanced models showed better discrimination than clinically reported FRAX scores in this DXA-tested population, but calibration assessment, prospective evaluation, and implementation workflow assessment are needed before clinical use.
\end{abstract}

\begin{keywords}
DXA, Osteoporosis \and Fracture risk assessment \and Statistical Methods \and Machine Learning \and General population studies
\end{keywords}

\section{Introduction}

Fragility fractures are a major consequence of low bone strength and remain a substantial cause of morbidity, mortality, loss of independence, and healthcare utilization in older adults \cite{LeBoff2022}. In the United States, more than 10 million adults aged 50 years and older are estimated to have osteoporosis, and the burden is expected to increase further as the population ages \cite{Wright2014, Gillespie2017}. Fractures associated with osteoporosis, particularly vertebral and hip fractures, are linked to hospitalization, prolonged rehabilitation, excess mortality, and major economic cost \cite{Brauer2009, Moreland2023, Burge2007, Lewiecki2018}.

Despite this burden, osteoporosis remains both underdiagnosed and undertreated. Recent studies suggest that a large proportion of individuals at high skeletal risk are never identified, even after a fracture event \cite{Desai2018, Choksi2023, Naso2025}. Improving fracture prevention, therefore, depends not only on treatment effectiveness but also on more accurate identification of individuals at near-term and mid-term risk.

The Fracture Risk Assessment Tool (FRAX) is the most commonly used clinical tool for fracture risk assessment \cite{Kanis2008}. Although broadly useful, FRAX has several well-recognized limitations. It includes only glucocorticoid exposure among medications affecting skeletal health, does not account for many other bone-adverse therapies, does not incorporate osteoporosis treatment history, and is primarily designed to estimate 10-year fracture probability rather than shorter-term risk \cite{Panday2014}. In addition, FRAX relies on femoral neck (FN) bone mineral density (BMD) when BMD is included and may not fully capture clinically relevant discordance across skeletal sites.

The increasing availability of longitudinal electronic health record (EHR) data offers an opportunity to improve fracture risk prediction beyond traditional risk calculators. EHR data can capture comorbidity burden, medication exposure, prior fracture history, and treatment patterns at scale. Machine learning (ML) methods may further exploit nonlinear relationships and interactions among these variables. However, the comparative value of ML relative to conventional survival modeling remains uncertain, and external validation across health systems has been limited \cite{Lehmann2024, Almog2020}.

In this retrospective clinical prediction study, we developed and externally validated fracture prediction models among adults aged 50 years or older who had clinically obtained DXA reports in 2 large US health care systems. We compared penalized Cox proportional hazards regression with 3 machine-learning survival approaches using prespecified feature sets derived from DXA reports and structured EHR data, and we benchmarked model discrimination against FRAX probabilities recorded in DXA reports. Because the intended-use population is patients undergoing DXA rather than the general population, we focused on post-DXA fracture risk stratification and on the transparency, transportability, and limitations of EHR-enhanced prediction.

\section{Materials and Methods}

\subsection{Study design and cohorts}

\begin{figure}[t]
\centering
\includegraphics[width=\textwidth]{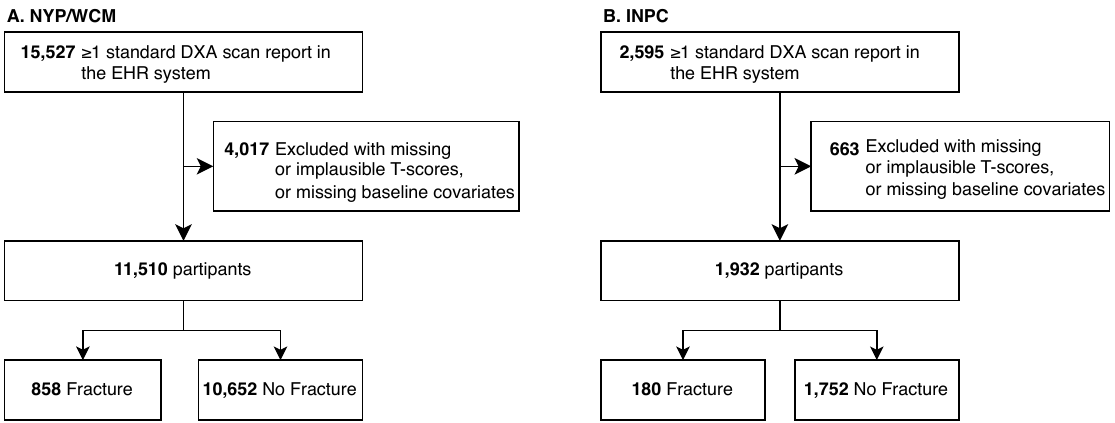}
\caption{\textbf{Participant selection flowchart for the development and external validation cohorts.}
Flowchart showing inclusion and exclusion of adults aged 50 years and older from the NewYork-Presbyterian/Weill Cornell Medical Center (NYP/WCM) development cohort and the Indiana Network for Patient Care (INPC) external validation cohort. Participants were required to have at least 1 standard dual-energy X-ray absorptiometry (DXA) report in the electronic health record. Individuals were excluded for missing or implausible T-scores or missing essential baseline covariates. FRAX was complete in the final NYP/WCM analytic cohort and was required in INPC to permit paired model-versus-FRAX external validation on the same individuals.}
\label{fig:workflow}
\end{figure}

We conducted a retrospective cohort study using EHR data from 2 healthcare systems (\textbf{Figure \ref{fig:workflow}}). The development cohort was derived from NewYork-Presbyterian/Weill Cornell Medical Center (NYP/WCM) and included adults aged 50 years and older who underwent standard DXA testing between January 1, 2012, and March 30, 2025. The external validation cohort was drawn from the Indiana Network for Patient Care (INPC), a statewide health information exchange maintained by the Indiana Health Information Exchange and made available for research through Regenstrief Institute and Indiana University infrastructure; this cohort included adults aged 50 years and older who underwent DXA testing between January 1, 2015, and April 1, 2025.

The intended-use population for the models was adults aged 50 years or older who had undergone clinically indicated DXA testing and had sufficient longitudinal EHR data for follow-up. The models were not designed for population-wide fracture screening among adults without DXA testing. This distinction is important because patients referred for DXA may differ from the general population in osteoporosis risk, comorbidity burden, health care use, and medication exposure.

Participants were required to have at least 1 standard DXA report in the EHR. Individuals were excluded for missing or implausible T-scores or missing essential baseline covariates. FRAX major osteoporotic fracture probability was extracted from the index DXA report and used only as a benchmark comparator, not as a predictor in the EHR/DXA models. In the NYP/WCM final analytic cohort, FRAX was available for all participants, so no NYP/WCM participants were excluded because of missing FRAX. Because the INPC cohort was used for external validation with paired comparison against FRAX, the INPC analytic cohort was restricted to participants with available FRAX values so that all models and FRAX could be evaluated on the same individuals. After applying these criteria, the final analytic sample included 11,510 participants in NYP/WCM and 1,932 participants in INPC.

The site-specific study data windows were January 1, 2012 to March 30, 2025 for NYP/WCM and January 1, 2015 to April 1, 2025 for INPC. Index DXA examinations, baseline covariates, and fracture outcome ascertainment were restricted to records available within these site-specific data windows. Follow-up for fracture outcomes continued from the index DXA date until the first qualifying incident fracture, the last recorded clinical encounter, or the end of the site-specific data window, whichever occurred first.

We used a complete-case analytic cohort for variables considered essential for model fitting, including valid follow-up time, fracture status, age, sex, BMI, and DXA-derived T-scores. T-scores greater than $+1$ were excluded because degenerative changes, including osteoarthritis, may artifactually elevate DXA measurements and bias fracture-risk estimates. Smoking and alcohol variables were retained as categorical EHR-derived status variables; unavailable or indeterminate values were encoded as "Unknown" rather than treated as missing. This approach preserved sample size and reflected how these fields would be available in routine EHR data, but it may introduce bias if unknown status is associated with health care use, documentation practices, or fracture risk.

The study was approved by the institutional review boards of Weill Cornell Medicine and Indiana University/Regenstrief Institute, with waiver of informed consent because the study used retrospective electronic health record data. Patient-level data were analyzed locally and were not shared across sites.

\subsubsection{Outcome definition}

The primary outcome was time from index DXA to first incident fragility fracture, identified from structured EHR data using ICD-10 code prefixes listed in \textbf{eTable \ref{tab:icd}}. Incident fragility fractures included forearm/wrist, hip/femur, lower leg/ankle/foot, shoulder/humerus, and spine/vertebral fractures; vertebral compression fractures and vertebral/spine fractures were combined into a single spine/vertebral category. Fractures due to major trauma (e.g., motor vehicle accidents) and those related to bone metastases were excluded when identifiable in structured EHR data. This outcome definition is not identical to the FRAX major osteoporotic fracture endpoint, which classically focuses on hip, clinical spine, forearm, and humerus fractures. Therefore, FRAX comparisons should be interpreted as benchmark comparisons against clinically reported FRAX probabilities rather than as direct validation of an identical endpoint.

For participants with an incident fracture, event time was defined as the interval from the index DXA date to the first qualifying fracture. For those without a fracture, follow-up time was defined as the interval from index DXA to the last recorded clinical encounter and treated as right-censored.

\subsubsection{Covariates}

Baseline covariates were obtained from structured EHR data and from DXA reports (\textbf{eTable \ref{tab:model_vars_matrix}}). Structured EHR variables included age, sex, body mass index (BMI), smoking status, alcohol use, prior fracture history, comorbidities, medications with negative skeletal potential (glucocorticoids, gonadotropin releasing hormone {[}GnRH{]} analogs, aromatase inhibitors, enzyme-inducing anticonvulsants, antiretrovirals, calcineurin inhibitors), and osteoporosis treatment (bisphosphonates, denosumab, teriparatide, abaloparatide, and romosozumab) exposures before the index DXA date. Baseline covariates were defined using information recorded before the index DXA date and within the site-specific study data window. For diagnosis-based comorbidities and prior medication exposure variables, all available structured records before the index DXA date were considered; no additional fixed maximum lookback interval was imposed. This approach was chosen because the selected diagnoses were generally chronic or clinically persistent risk factors, and the medication variables represented prior exposure before DXA rather than post-index treatment. Prior fracture was defined using qualifying fracture codes before the index DXA date. Incident fracture outcomes were restricted to qualifying fracture codes after index DXA.

\subsubsection{DXA report extraction}

DXA-derived variables were extracted from radiology reports using a custom rule-based natural language processing pipeline implemented with regular expressions. The algorithm identified the scan date and extracted T-scores for the lumbar spine (LS), femoral neck (FN), and total hip (TH), as well as FRAX 10-year major osteoporotic fracture probability when reported. Details of the extraction pipeline are provided in \textbf{eSection \ref{sec:nlp}}.

In addition to site-specific T-scores, we derived the minimum T-score across available central DXA sites, defined as the lowest value among LS, FN, and TH T-scores for each participant. Trabecular bone score and forearm measures were not analyzed because of substantial missingness. To reduce potential distortion from degenerative change or artifact, T-scores greater than +1 were excluded. For benchmarking, FRAX's major osteoporotic fracture probability was extracted directly from DXA reports and treated as a fixed baseline risk score.

\subsection{Statistical analysis}\label{statistical-analysis}

\subsubsection{Model settings}\label{model-settings}

We prespecified 2 feature sets (\textbf{eTable \ref{tab:model_vars_matrix}}). \modela was designed as an EHR-enhanced fracture prediction model and included demographics, lifestyle factors, the minimum central DXA T-score, comorbidity burden, medications with negative skeletal potential, and osteoporosis treatment history. In \modela, individual comorbidities and medication classes were first encoded as binary indicators and then summarized as prespecified counts to limit model complexity relative to the number of fracture events. The comorbidity count included type 1 diabetes, type 2 diabetes, hyperparathyroidism, rheumatoid arthritis, systemic lupus erythematosus, ankylosing spondylitis, celiac disease, ulcerative colitis, Crohn disease, hypogonadism, premature menopause, eating disorder, Cushing syndrome, chronic kidney disease, asthma, chronic obstructive pulmonary disease, cystic fibrosis, cirrhosis, multiple myeloma, stroke, Parkinson disease, and Alzheimer disease. The adverse skeletal medication count included systemic glucocorticoids, GnRH analogs or antagonists, aromatase inhibitors, enzyme-inducing antiepileptic drugs, antiretroviral drugs, and calcineurin inhibitors. The osteoporosis treatment count included bisphosphonates, denosumab, teriparatide, abaloparatide, and romosozumab. \modelb was designed to more closely mirror the clinical structure of FRAX while retaining a survival-modeling framework. It included demographics, lifestyle factors, FN T-score, rheumatoid arthritis, and systemic steroid exposure.

For Cox models, continuous predictors were modeled as linear terms after standardization; restricted cubic splines or other nonlinear transformations were not used. Hazard ratios from the final full-cohort Cox models were converted from standardized coefficients to clinically interpretable contrasts: age per 10-year increase, BMI per 5-kg/m2 increase, T-scores per 1-unit higher value, binary predictors as yes versus no, and burden variables per 1 additional condition or medication group. The tree-based machine-learning models could capture nonlinear associations and interactions through recursive partitioning or boosting.

\subsubsection{Model development}\label{model-development}

We fit penalized Cox proportional hazards models for both \modela and \modelb using time to first incident fragility fracture as the outcome \cite{cox1972regression, harrell1996multivariable}. A fixed penalization parameter of 0.01 was used for all Cox models.

The survival models produced relative risk scores for ranking participants by predicted fracture risk after index DXA. Model discrimination was evaluated over the observed follow-up period and at prespecified time horizons of 1, 2, and 5 years. These model scores were not recalibrated to estimate 10-year absolute fracture probability. FRAX major osteoporotic fracture probability, extracted from DXA reports, represents a 10-year clinical risk estimate. Therefore, comparisons with FRAX emphasize discrimination and risk ranking rather than equivalence of absolute risk horizons.

We also developed 3 ML survival models using the same predictor sets: random survival forest (RSF) \cite{ishwaran2008random, jaeger2019oblique}, gradient-boosting survival analysis (GBS) \cite{hothorn2006survival, chen2013gradient}, and XGBoost survival with a Cox objective \cite{chen2016xgboost}. RSF was implemented with 800 trees, square-root feature subsampling, and minimum split and terminal node sizes of 10 and 5, respectively. GBS used 1,500 boosting iterations, a learning rate of 0.05, a maximum tree depth of 3, minimum split and leaf sizes of 10 and 5, and a subsampling fraction of 0.8. XGBoost survival models were trained with a Cox partial-likelihood objective, a learning rate of 0.05, a maximum depth of 3, a minimum child weight of 1.0, subsampling of 0.8, column subsampling of 0.8, L2 regularization of 1.0, and L1 regularization of 0.0.

\subsubsection{Internal and external validation}\label{internal-and-external-validation}

Internal validation was performed in the NYP/WCM cohort using Monte Carlo repeated random train-test splitting. In each of 20 iterations, 80\% of the cohort was assigned to the training set and 20\% to the test set. All candidate models and FRAX were evaluated on the same held-out test set within each iteration, allowing directly paired comparisons.

For external validation, final models were trained on the full NYP/WCM cohort (100\% of WCM data) and then applied directly to the INPC cohort without refitting. FRAX was evaluated in the same external cohort as a benchmark comparator.

\subsubsection{Performance assessment}\label{performance-assessment}

Model discrimination was assessed using Harrell's concordance index (C-index) \cite{harrell1996multivariable}, Uno's inverse-probability-of-censoring-weighted (IPCW) C-index \cite{uno2011c}, and time-dependent AUC \cite{heagerty2000time}. Harrell's C-index summarized overall rank discrimination across the entire follow-up period. Uno's IPCW C-index was calculated at 1, 2, and 5 years to provide time-specific discrimination while accounting for censoring. Time-dependent AUCs were also calculated at 1, 2, and 5 years. To ensure stable estimation, time-dependent analyses were restricted to follow-up of less than 6 years.

For internal validation, performance was summarized across the 20 repeated splits as mean and 95\% confidence interval (CI). Pairwise comparisons were conducted using paired t-tests and Wilcoxon signed-rank tests. For external validation, 1,000 bootstrap samples were used to estimate uncertainty and compare differences in discrimination metrics.

\subsubsection{Statistical software}\label{statistical-software}

Analyses were performed using Python version 3.11.14 with pandas, NumPy, SciPy, scikit-learn, scikit-survival, lifelines, and XGBoost, together with related scientific computing packages.

\section{Results}\label{results}

\subsection{Cohort characteristics}\label{cohort-characteristics}

The NYP/WCM cohort included 11,510 adults, of whom 858 experienced an incident fragility fracture. The INPC cohort included 1,932 participants, of whom 180 experienced a fragility fracture. \textbf{Table \ref{tab:fracture}} shows baseline characteristics. In NYP/WCM, participants who subsequently fractured were older and more likely to be current or former smokers and to have a history of asthma, chronic kidney disease, chronic obstructive pulmonary disease, multiple myeloma, Parkinson's disease, rheumatoid arthritis, stroke, Alzheimer's disease, and type 2 diabetes mellitus. They also had lower LS, FN, and TH T-scores, were more likely to have a prior fracture history, and more frequently had prior exposure to bisphosphonates, abaloparatide, and teriparatide. In INPC, participants with incident fractures had higher BMI, were more likely to be male, and were more likely to have a prior fracture history. Asthma, rheumatoid arthritis, type 2 diabetes mellitus, and current alcohol use were also more common among those who fractured. Median follow-up was 35.4 months in NYP/WCM and 32.3 months in INPC (\textbf{eFigure \ref{fig:followup_gap}}). In NYP/WCM, vertebral/spine fractures were the most common incident fracture type ($n=376$), followed by forearm/wrist ($n=111$) and hip/femur fractures ($n=98$).
{
% \tiny
\begin{landscape}
\input{tables/tab_fracture}
\end{landscape}
}

\subsection{Internal validation performance}\label{internal-validation-performance}

\subsubsection{Comparison with FRAX}\label{comparison-with-frax}

\begin{figure}[thp]
    \centering
    \includegraphics[width=.6\textwidth]{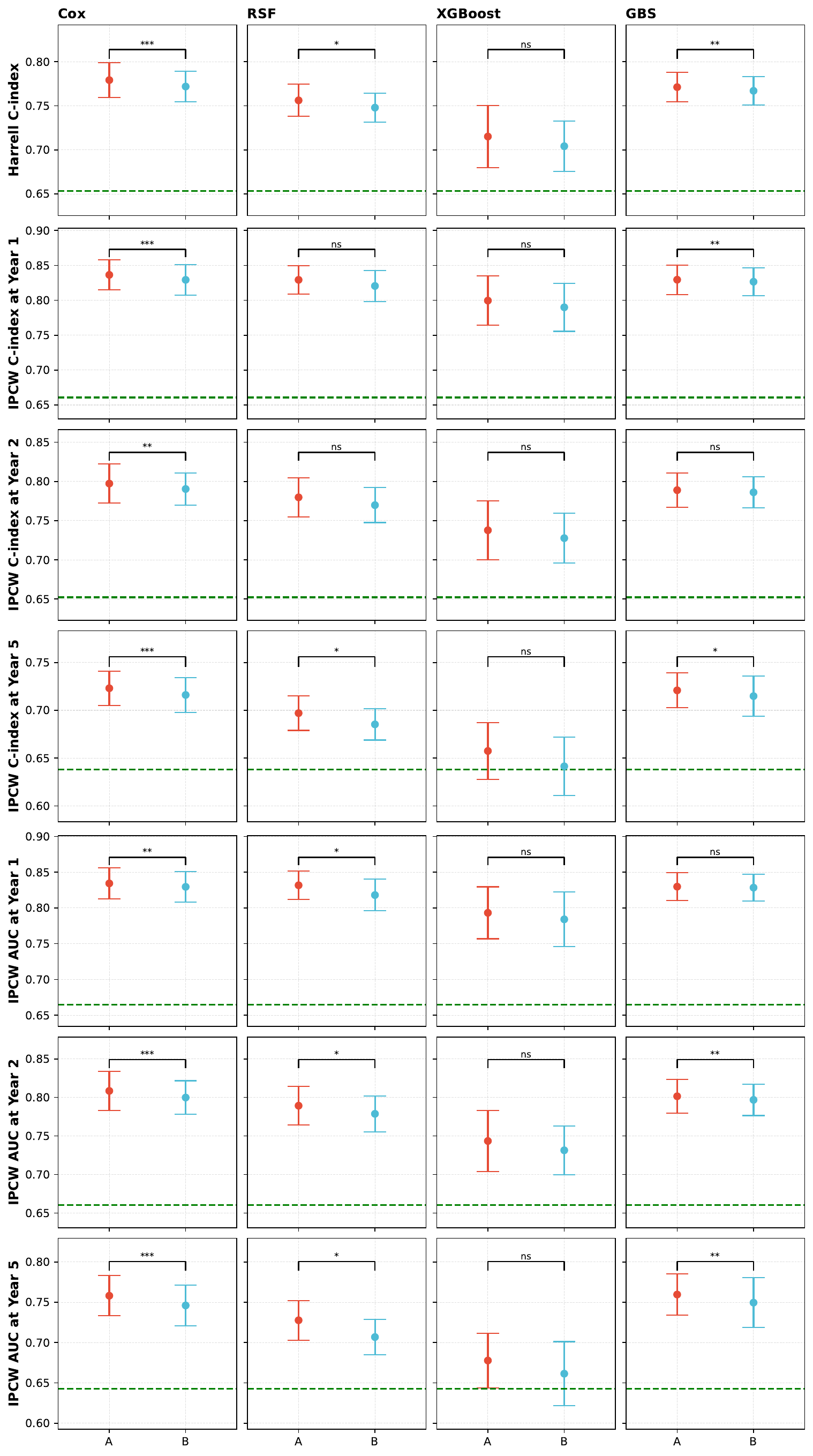}
    \caption{\textbf{Discrimination of Cox, machine learning survival models, and FRAX in internal validation.} 
    Model performance in the NYP/WCM cohort across 20 repeated random 80/20 train-test splits. Shown are Harrell's C-index, Uno's inverse-probability-of-censoring-weighted (IPCW) C-index at 1, 2, and 5 years, and time-dependent IPCW area under the receiver operating characteristic curve (AUC) at 1, 2, and 5 years for \modela and \modelb. Points represent mean values, and error bars indicate 95\% confidence intervals across repeated splits. Green dashed lines indicate FRAX's performance. \modela incorporated comprehensive clinical risk factors, DXA-derived minimum T-score, medication exposures with negative skeletal potential, and osteoporosis treatment history; \modelb was designed to more closely parallel the FRAX clinical framework.
    }
    \label{fig:internal_barplots}
\end{figure}

In internal validation, models that integrated DXA and structured EHR variables substantially outperformed FRAX across all discrimination metrics (\textbf{Figure \ref{fig:internal_barplots}}). The best-performing model overall was the Cox model in \modela, which achieved a mean Harrell's C-index of 0.779 (95\% CI, 0.770--0.788). This was modestly higher than the Cox model in \modelb and markedly better than FRAX, which achieved a Harrell's C-index of 0.653 (95\% CI, 0.646--0.660).

Time-specific discrimination showed the same pattern. For the Cox model in \modela, Uno IPCW C-indexes were 0.836, 0.797, and 0.723 at 1, 2, and 5 years, respectively, compared with 0.829, 0.790, and 0.716 for the Cox model in \modelb and 0.661, 0.652, and 0.638 for FRAX. Time-dependent AUCs for the Cox model in \modela were 0.834, 0.808, and 0.758 at 1, 2, and 5 years, respectively, compared with 0.794, 0.776, and 0.743 for FRAX.

\subsubsection{Comparison across model families}\label{comparison-across-model-families}

Among the ML approaches, GBS showed the strongest discrimination, followed by RSF and XGBoost. For Harrell's C-index, GBS achieved 0.771 in \modela and 0.767 in \modelb, compared with 0.756 and 0.748 for RSF and 0.716 and 0.706 for XGBoost.

Time-specific performance showed similar patterns. GBS achieved Uno IPCW C-indexes of 0.830, 0.789, and 0.721 at 1, 2, and 5 years and time-dependent AUCs of 0.830, 0.801, 0.767, 0.745, and 0.760 at 1 through 5 years. RSF showed intermediate performance, whereas XGBoost consistently performed worst among the EHR-enhanced models.

Despite GBS's strong performance, the Cox model in \modela remained significantly superior to all ML approaches in paired internal comparisons. Compared with the Cox model in \modela, mean Harrell's C-index was lower for GBS (0.771 versus 0.779; paired t-test $P < .01$; Wilcoxon $P < .01$), RSF (0.756 versus 0.779; paired t-test $P < .01$; Wilcoxon $P < .01$), and XGBoost survival (0.716 versus 0.779; paired t-test $P < .01$; Wilcoxon $P < .01$).

\subsection{External validation performance}\label{external-validation-performance}

\subsection{Comparison with FRAX}\label{comparison-with-frax-1}

\begin{figure}[pht]
    \centering
    \includegraphics[width=.65\linewidth]{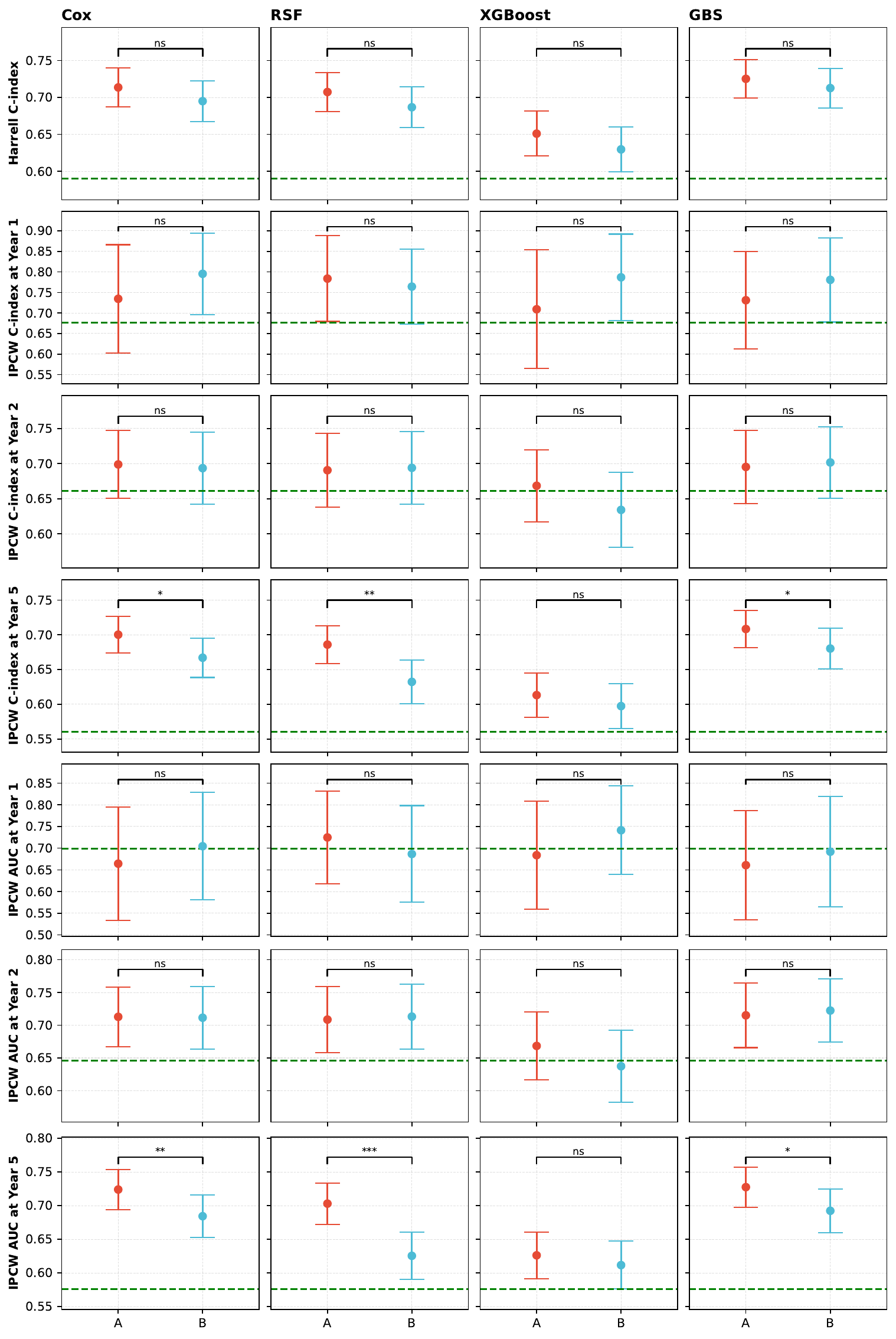}
    \caption{\textbf{Discrimination of Cox, machine learning survival models, and FRAX in external validation.} Model performance after training on the full NYP/WCM cohort and direct application to the independent INPC cohort without refitting. Shown are Harrell's C-index, Uno's inverse-probability-of-censoring-weighted (IPCW) C-index at 1, 2, and 5 years, and time-dependent IPCW area under the receiver operating characteristic curve (AUC) at 1, 2, and 5 years for \modela and \modelb. Points represent mean values, and error bars indicate bootstrap-based 95\% confidence intervals from 1,000 external-validation bootstrap samples. Green dashed lines indicate FRAX's performance. \modela incorporated comprehensive clinical risk factors, DXA-derived minimum T-score, medication exposures with negative skeletal potential, and osteoporosis treatment history; \modelb was designed to more closely parallel the FRAX clinical framework.}
    \label{fig:external_barplots}
\end{figure}

Final models trained in the full NYP/WCM cohort were then applied directly to the INPC cohort without refitting (\textbf{Figure \ref{fig:external_barplots}}). In external validation, the Cox model in \modela retained good discrimination, with a Harrell's C-index of 0.714 (95\% CI, 0.662--0.765), compared with 0.695 (95\% CI, 0.641--0.749) for the Cox model in \modelb and 0.590 (95\% CI, 0.544--0.637) for FRAX.

For the Cox model in \modela, Uno IPCW C-indexes were 0.735 (95\% CI, 0.476--0.993), 0.699 (95\% CI, 0.604--0.794), and 0.700 (95\% CI, 0.648--0.752) at 1, 2, and 5 years, respectively. The corresponding values for the Cox model in \modelb were 0.795, 0.694, and 0.667, whereas FRAX achieved 0.676, 0.661, and 0.561. Bootstrap testing showed that the Cox model in \modela significantly outperformed both the Cox model in \modelb and FRAX in overall discrimination.

\subsubsection{Comparison across model families}\label{comparison-across-model-families-1}

Although the Cox model in \modela showed the best internal performance, GBS achieved the highest overall discrimination in external validation. In the FRAX-available INPC subset, Harrell's C-index was 0.725 (95\% CI, 0.675--0.776) for GBS, followed by 0.714 (95\% CI, 0.662--0.765) for Cox regression, 0.707 (95\% CI, 0.655--0.759) for RSF, and 0.651 for XGBoost. FRAX remained substantially lower at 0.590.

Across ML models, \modela again outperformed \modelb, although the between-setting differences were not statistically significant on bootstrap testing. The same ranking was generally observed for time-specific Uno IPCW C-indexes and time-dependent AUCs. For GBS, Uno IPCW C-indexes were 0.731, 0.695, and 0.708 at 1, 2, and 5 years, with corresponding AUCs of 0.661, 0.715, and 0.727.

\subsection{Model interpretation}\label{model-interpretation}

\begin{figure}[t]
\centering
\begin{subfigure}[b]{0.47\textwidth}
\centering
\includegraphics[width=\linewidth]{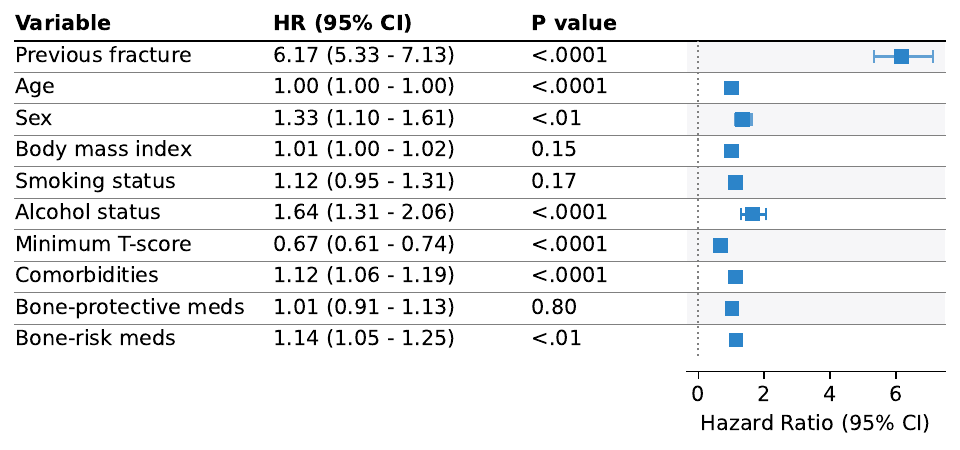}
\caption{\modela}
\label{fig:hr_wcm_a}
\end{subfigure}\hfill
\begin{subfigure}[b]{0.47\textwidth}
\centering
\includegraphics[width=\linewidth]{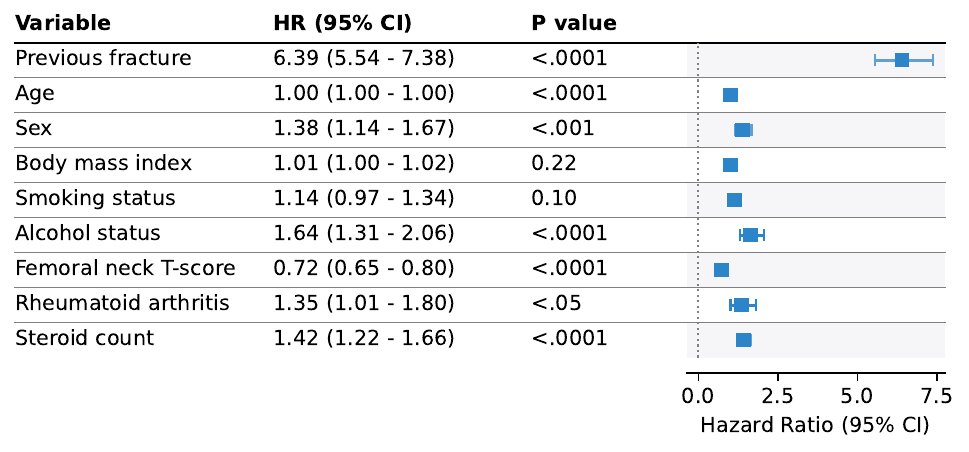}
\caption{\modelb}
\label{fig:hr_wcm_b}
\end{subfigure}
\caption{\textbf{Adjusted hazard ratios for incident fragility fracture in Cox models for \modela and \modelb.}
Forest plots show adjusted hazard ratios (HRs) and 95\% confidence intervals from the final Cox proportional hazards models fit in the NYP/WCM cohort. \modela included demographics, lifestyle factors, DXA-derived minimum T-score, comorbidity burden, medication exposures with negative skeletal potential, and osteoporosis treatment history. \modelb included demographics, lifestyle factors, femoral neck T-score, rheumatoid arthritis, and systemic steroid exposure. HRs greater than 1 indicate higher fracture risk; HRs less than 1 indicate lower fracture risk.
}
\label{fig:hazard_ratio}
\end{figure}

\textbf{Figure \ref{fig:hazard_ratio}} shows that prior fracture was the strongest predictor in both Cox settings (\modela: HR, 6.16; 95\% CI, 5.29-7.18; \modelb: HR, 6.40; 95\% CI, 5.52-7.41). Older age was also associated with higher fracture risk, with HRs of 1.38 per 10-year increase in both settings. Higher DXA T-score was protective (minimum T-score in \modela: HR, 0.67; 95\% CI, 0.61-0.75; femoral neck T-score in \modelb: HR, 0.72; 95\% CI, 0.64-0.80). In \modela, EHR-coded alcohol status, comorbidity burden, adverse skeletal medication burden, and male sex were associated with higher risk, whereas BMI, smoking status, and osteoporosis treatment burden were not statistically significant. In \modelb, EHR-coded alcohol status, systemic steroid exposure, rheumatoid arthritis, and male sex were associated with higher risk, whereas BMI and smoking status were not statistically significant.

\section{Discussion}\label{discussion}

In this retrospective development and external validation study of adults aged 50 years or older with clinically obtained DXA reports, models combining DXA-derived T-scores with structured EHR information showed better discrimination for incident fragility fracture than FRAX probabilities recorded in DXA reports. The strongest internal performance was observed for a parsimonious penalized Cox model using the minimum central DXA T-score, prior fracture, demographic and lifestyle variables, and aggregated comorbidity and medication measures. In external validation, gradient-boosting survival had the highest discrimination, while the Cox model remained competitive and more interpretable. These findings support the potential value of routinely collected EHR and DXA data for post-DXA risk stratification, but they should be interpreted as evidence of improved risk ranking rather than proof that the models are ready for clinical implementation.

The comparison with FRAX requires caution. FRAX probabilities were extracted from DXA reports rather than recalculated from raw FRAX inputs, and FRAX may have been obtained in clinical contexts where clinicians already perceived elevated fracture risk. Such patients may have subsequently received counseling or pharmacologic treatment, potentially attenuating observed fracture incidence. In NYP/WCM, FRAX was available for all participants in the final analytic cohort; in INPC, the external validation cohort was restricted to participants with available FRAX values to permit paired model-versus-FRAX evaluation on the same individuals. This restriction may limit generalizability to DXA-tested patients without documented FRAX values. In addition, FRAX estimates 10-year major osteoporotic fracture probability and incorporates mortality assumptions, whereas our primary models were evaluated as time-to-event risk scores over observed follow-up and at 1-, 2-, and 5-year horizons. Finally, our EHR-defined fragility fracture outcome was broader than the classic FRAX major osteoporotic fracture endpoint. For these reasons, FRAX should be viewed as a clinically relevant benchmark rather than a perfectly matched comparator.

These findings are clinically important for several reasons. First, they show that substantial fracture risk information is already present in routinely collected structured EHR data and can be leveraged without specialized testing beyond DXA. Second, they suggest that prediction of shorter-term fracture risk can be materially improved relative to FRAX, which is primarily designed for 10-year probability estimation~\cite{Dickens2024-hn}. Third, they support a pragmatic approach to implementation: an interpretable Cox model may be sufficient in many settings, whereas ML may provide incremental gains in some externally transported contexts.

One notable finding was the consistent value of using the minimum T-score across the LS, FN, and TH. FRAX relies on FN BMD when BMD is incorporated, yet discordance among skeletal sites is common in clinical practice \cite{Leslie2014}. A model based on the lowest DXA T-score may better capture skeletal fragility, particularly when LS values are substantially lower than FN values or when vertebral risk is of interest. Our results suggest that restricting prediction to the FN alone may underuse clinically informative DXA data.

Prior fracture, alcohol use and smoking status, and sex were the most influential predictors across all model families, which is biologically and clinically plausible. DXA-derived T-scores remained central to model performance, but they were not sufficient on their own. Comorbidity burden, medication exposures with negative skeletal potential, and prior osteoporosis treatment all contributed additional information.

The relative performance of Cox and ML methods warrants careful interpretation. In internal validation, Cox regression performed best and significantly outperformed all 3 ML approaches. This is consistent with prior studies showing that more flexible ML algorithms do not consistently outperform traditional regression approaches in clinical prediction modeling \cite{christodoulou2019systematic, nusinovici2020logistic}. At the same time, external validation told a more nuanced story: gradient-boosting survival showed the highest overall discrimination in the independent cohort. This pattern suggests that although Cox regression provided the most robust model during development, some ML methods may capture transportable nonlinearities or interactions not fully represented by the Cox specification.

Our findings also align with the broader fracture prediction literature. Some studies have reported stronger performance of traditional regression-based models, whereas others have found an advantage for ML \cite{Lehmann2024, christodoulou2019systematic, li2023sexspecific, jaiswal2025hrpqct, devries2021subsequent, kruse2017hip}. Such differences likely reflect variation in cohort composition, predictor availability, fracture definitions, time horizons, and the degree of heterogeneity between development and validation populations. Importantly, our results do not support a simple narrative that ML is uniformly better than conventional methods. Rather, they suggest that model choice should be guided by the clinical use case, the need for interpretability, and the expected deployment setting.

This study has several strengths. It used 2 large real-world cohorts from distinct healthcare systems, included both internal and external validation, compared traditional and machine-learning survival methods using the same feature definitions, and benchmarked performance against FRAX. The DXA extraction pipeline enabled integration of clinically relevant radiology report data into structured models, and the use of time-to-event methods yielded clinically interpretable, shorter-term discrimination estimates.

Generalizability is constrained by the DXA-tested source population and the demographic composition of both cohorts, which predominantly included women and White patients. Patients without DXA access or regular health care contact may have different risk profiles and data completeness. Documentation of lifestyle factors, diagnoses, medications, and fractures may also vary across health systems. Additional validation is needed in men, racially and ethnically diverse populations, and systems with different DXA referral and EHR coding practices.

Several limitations should also be acknowledged. First, the study was retrospective and limited to individuals who underwent DXA and had sufficient follow-up, potentially limiting generalizability to broader populations without DXA access. Second, comorbidities and medication exposures were derived from structured EHR data and are therefore susceptible to incomplete capture, coding variation, and exposure misclassification. Third, some established fracture risk factors, including a history of falls and functional status, were not reliably available. Fourth, vertebral fractures may have been under ascertained because many are clinically silent. In addition, longer-horizon estimates are inherently less stable as censoring accumulates and the number at risk decreases. Finally, the external benchmark comparison with FRAX was performed in the FRAX-available subset of INPC, thereby improving comparability but potentially limiting generalizability.

Excluding individuals with missing essential predictors or implausible T-scores may have introduced selection bias and reduced generalizability. Patients with incomplete DXA reports, missing BMI, or sparse EHR data may differ systematically from those retained in the analytic cohort. We retained unknown smoking and alcohol status as explicit categories, but other variables were handled using complete-case criteria. Future implementation would require a prespecified approach for unavailable predictor values at the point of care, such as validated imputation strategies or model versions using reduced predictor sets.

The sample size was determined by available retrospective EHR and DXA data rather than by a prospective calculation. Although the development cohort included 858 fracture events and the models used prespecified, relatively compact feature sets, the external validation cohort included 180 fracture events, limiting precision for subgroup, calibration, and time-specific analyses. We therefore interpret subgroup and external calibration results cautiously and view this study as a step toward, rather than a substitute for, prospective implementation evaluation.

In adults aged 50 years or older with clinically obtained DXA reports, EHR- and DXA-enhanced survival models showed better fracture-risk discrimination than clinically reported FRAX probabilities in both internal and external validation. The results support further development of transparent post-DXA risk stratification tools that use routinely collected clinical data. Before clinical deployment, future studies should evaluate calibration, competing-risk approaches, missing-data workflows, subgroup performance, and prospective clinical utility across more diverse health systems.

\textbf{Conflict of Interest Disclosures}

None reported.

\textbf{Funding}

This work is supported in part by funds from the U.S. National Institutes of Health (R01CA289249 and P30AR072581) and U.S. National Science Foundation (NSF SCH 2306556 and NSF CAREER 2145640).

\textbf{Role of the Funder}

The funders had no role in the design and conduct of the study; collection, management, analysis, and interpretation of the data; preparation, review, or approval of the manuscript; or decision to submit the manuscript for publication.

\textbf{Data availability}

The individual-level data underlying this article cannot be shared publicly because they contain protected health information from electronic health records and are subject to institutional data-use restrictions. Deidentified aggregate results and analytic code may be made available from the corresponding author upon reasonable request, subject to institutional review, data-use agreements, and applicable regulatory approvals.

\textbf{Ethics approval}

This study was approved by the Institutional Review Boards of Weill Cornell Medicine and Indiana University/Regenstrief Institute.

\textbf{Patient consent}

The requirement for informed consent was waived by the Institutional Review Boards of Weill Cornell Medicine and Indiana University/Regenstrief Institute because this retrospective study used electronic health record data.

\textbf{Clinical trial registration}

Not applicable.

\textbf{Author contributions statement}

J.Q., H.D., J.B., Y.P., and Y.L. conceived and designed the study. J.B., Y.P., and Y.L. provided data resources and supervised the study. J.Q., H.D., K.Y., and E.I. reviewed the methods and interpreted the results. J.Q., K.Y., and He.D. collected and processed the data, performed the internal validation, analyzed the data, organized the results, reviewed the findings, and drafted the manuscript. H.D., X.H., E.I., and J.B. reviewed the methods and performed the external validation. E.I. and Y.L. provided clinical expertise and data interpretation, conducted the literature review, analyzed the results, and contributed to writing and revising the manuscript. All authors reviewed and approved the final manuscript.

\setlength{\bibsep}{3pt plus 0.3ex}
\bibliographystyle{unsrtnat}
\bibliography{reference}
%\printbibliography

\newpage
\appendix

\setcounter{table}{0}
\renewcommand\tablename{eTable}
\crefalias{table}{etable}

\setcounter{figure}{0}
\renewcommand\figurename{eFigure}
\crefalias{figure}{efigure}

\crefalias{section}{esection}

\part*{Supplementary materials}
\label{sec:appendix}

\begin{table}[h!]
\centering
\caption{\textbf{Distribution of incident fragility fracture categories in the NYP/WCM and INPC cohorts.} Fracture categories were defined using ICD-10 code prefixes recorded after the index DXA examination. Vertebral compression fractures and vertebral/spine fractures were combined into a single spine/vertebral category. Counts and percentages are shown for each cohort.
Fragility fractures due to major trauma or bone metastases were excluded.}
\label{tab:icd}
\renewcommand{\arraystretch}{1.1}
\small
\input{tables/tab_fracture_cat}
\end{table}

\clearpage

\begin{table}[h!]
\centering
\caption{\textbf{Variables included in the 2 prespecified fracture prediction settings.}
\modela represents the expanded EHR-enhanced model, including comprehensive clinical risk factors, DXA-derived minimum T-score, medication exposures with negative skeletal potential, and osteoporosis treatment history. 
\modelb represents the FRAX-aligned model specification, incorporating femoral neck T-score and selected clinical risk factors analogous to the FRAX framework.
BMI = body mass index; $T_{min}$ = minimum T-score across lumbar spine, femoral neck, and total hip; RA = rheumatoid arthritis.
}
\renewcommand{\arraystretch}{1.1}
\label{tab:model_vars_matrix}
\small
\input{tables/tab_variables}
\end{table}

\clearpage

\begin{table}[h!]
\centering
\caption{\textbf{Top predictors ranked by mean permutation importance across 4 survival model families in internal validation.} The 5 highest-ranked predictors are shown for Cox regression, XGBoost survival, random survival forest (RSF), and gradient-boosting survival (GBS) in \modela and \modelb. Importance values represent the mean reduction in model discrimination after permutation of each feature across repeated internal-validation splits; larger positive values indicate greater contribution to model performance.
RSF = random survival forest; GBS = gradient-boosting survival; minimum T-score = lowest value across available lumbar spine, femoral neck, and total hip T-scores; bone-protective medications = osteoporosis treatment burden.}
\label{tab:perm_top5_internal_4models}
\scriptsize
\input{tables/tab_predicator}
\end{table}

\clearpage

\begin{figure}[h!]
\centering
\includegraphics[width=\linewidth]{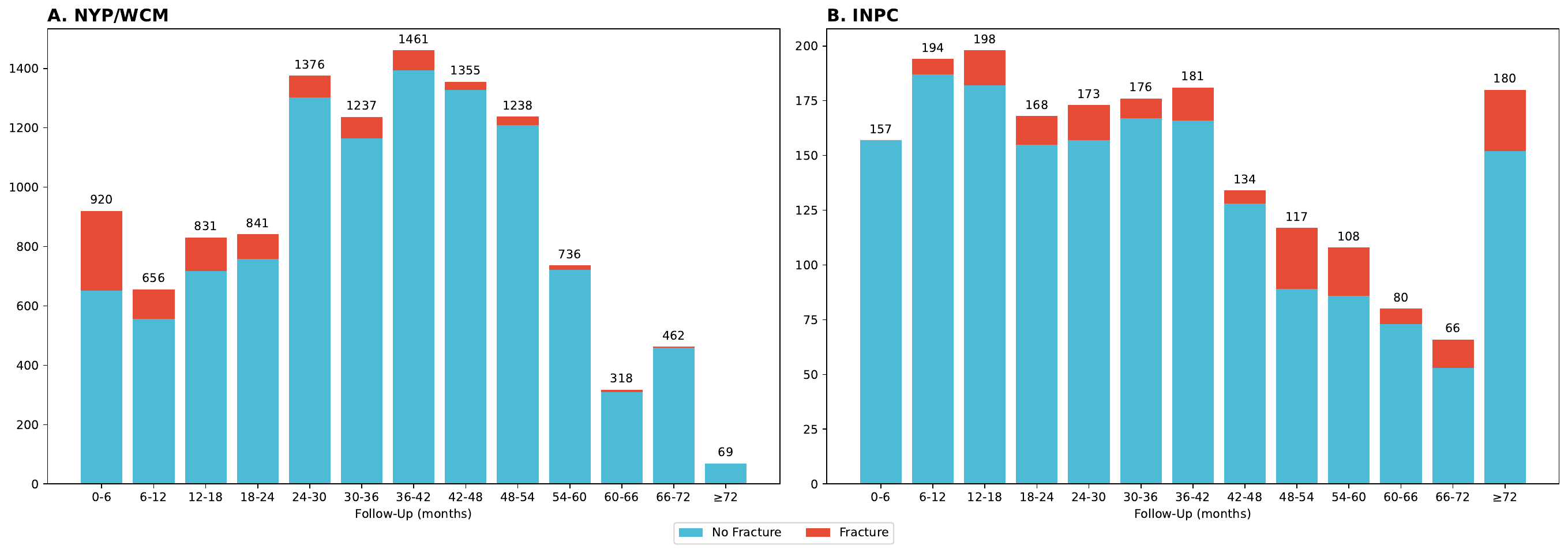}
\caption{\textbf{Distribution of follow-up time by site.} Stacked bar plots show the number of participants within each follow-up interval, stratified by incident fragility fracture status, in the NYP/WCM development cohort and the INPC external validation cohort. Follow-up time was defined as the interval from the index DXA date to the first qualifying fragility fracture or the last recorded clinical encounter for censored participants.}
\label{fig:followup_gap}
\end{figure}

\clearpage

\input{suppl}
\end{document}

%% file: tables/tab_fracture.tex
% --- add this once (preamble or right before the table) ---
\newcommand{\blankcell}{\hspace{0pt}}

\begin{longtblr}[
    theme=thesis,
    caption = {\textbf{Baseline characteristics of the NYP/WCM development cohort and INPC external validation cohort, stratified by incident fragility fracture.} Participant characteristics are shown for the overall cohort and according to whether an incident fragility fracture occurred during follow-up. Continuous variables are presented as mean (SD), and categorical variables are presented as number (\%).},
    label = {tab:fracture},
    remark{Note} = {Values are mean $\pm$ standard deviation or number (percentage). Percentages may not sum to 100 due to rounding.},
    note{a} = {P values were calculated by Welch's 2-sample t test for continuous variables and Pearson’s chi-square test for categorical variables.}
]{
    width = \linewidth,
    colspec={@{}lXlllrlllr@{}},
    rowhead = 2,
    hlines = {2-10}{stone},
    hline{1,3,Z} = {black},
    hline{4,5,8,14,18,22,26,27,31,54,60} = {stone},
    hline{2} = {3-5}{black},
    hline{2} = {6}{r,black},
    hline{2} = {7-10}{black},
    cells = {font = \scriptsize},
}
%
% -------------------------------------------------
% Header
% -------------------------------------------------
\SetCell[r=2,c=2]{l}{\textbf{Characteristic}}
  & \blankcell
  & \SetCell[c=4]{c}{\textbf{NYP/WCM Participants, No. (\%)}}
  & & &
  & \SetCell[c=4]{c}{\textbf{INPC Participants, No. (\%)}}
  & & & \\
& &
{\textbf{All}\\ (N=11510)} &
{\textbf{Fracture}\\ (n=858)} &
{\textbf{No Fracture}\\ (n=10652)} &
{\textbf{P value}\TblrNote{a}} &
{\textbf{All}\\ (N=1932)} &
{\textbf{Fracture}\\ (n=180)} &
{\textbf{No Fracture}\\ (n=1752)} &
{\textbf{P value}} \\
%
% -----------------------------
% Core demographics / DXA
% -----------------------------
\SetCell[c=2]{l}Age, mean (SD), y &&
69.37 (8.77) & 73.06 (8.87) & 69.07 (8.69) & $<$0.001 &
67.46 (7.95) & 68.09 (9.09) & 67.39 (7.82) & 0.322 \\
\SetCell[c=2]{l}Body mass index, mean (SD), kg/m$^2$ &&
25.38 (5.10) & 25.50 (5.30) & 25.37 (5.08) & 0.470 &
29.31 (7.04) & 30.45 (7.68) & 29.18 (6.96) & 0.049 \\
\SetCell[c=2]{l}Sex && & & & $<$0.001 & & & & 0.016 \\
& Female
  & 10515 (91.4) & 738 (86.0) & 9777 (91.8) & \blankcell
  & 1829 (94.7) & 163 (90.6) & 1666 (95.1) & \blankcell \\
& Male
  & 995 (8.6) & 120 (14.0) & 875 (8.2) & \blankcell
  & 103 (5.3) & 17 (9.4) & 86 (4.9) & \blankcell \\
\SetCell[c=2]{l}Race && & & & \blankcell & & & & 0.232 \\
& White
  & 7468 (64.9) & 626 (72.9) & 6842 (64.2) & \blankcell
  & 1356 (70.2) & 139 (77.2) & 1217 (69.5) & \blankcell \\
& Black
  & 943 (8.2) & 52 (6.1) & 891 (8.4) & \blankcell
  & 281 (14.5) & 22 (12.2) & 259 (14.8) & \blankcell \\
& Asian
  & 762 (6.6) & 44 (5.1) & 718 (6.7) & \blankcell
  & 5 (0.3) & 0 (0.0) & 5 (0.3) & \blankcell \\
& American Indian%/Alaska Native or Pacific Islander
  & 180 (1.6) & 9 (1.0) & 171 (1.6) & \blankcell
  & 6 (0.3) & 0 (0.0) & 6 (0.3) & \blankcell \\
& Others/No Matching Concept
  & 2157 (18.7) & 127 (14.8) & 2030 (19.1) & \blankcell
  & 284 (14.7) & 19 (10.6) & 265 (15.1) & \blankcell \\
\SetCell[c=2]{l}Ethnicity && & & & \blankcell & & & & 0.817 \\
& Not Hispanic or Latino
  & 8562 (74.4) & 682 (79.5) & 7880 (74.0) & \blankcell
  & 1891 (97.9) & 177 (98.3) & 1714 (97.8) & \blankcell \\
& Hispanic or Latino
  & 1070 (9.3) & 72 (8.4) & 998 (9.4) & \blankcell
  & 38 (2.0) & 3 (1.7) & 35 (2.0) & \blankcell \\
& No matching concept
  & 1878 (16.3) & 104 (12.1) & 1774 (16.7) & \blankcell
  & 3 (0.2) & 0 (0.0) & 3 (0.2) & \blankcell \\
%
% -----------------------------
% Smoking status
% -----------------------------
\SetCell[c=2]{l}Smoking status && & & & $<$0.001 & & & & 0.073 \\
& Never smoker
  & 9693 (84.2) & 690 (80.4) & 9003 (84.5)
  & \blankcell & 1416 (73.3) & 136 (75.6) & 1280 (73.1) & \blankcell \\
& Current/Former smoker
  & (5.4) & 70(8.2) & 546(5.1)
  & \blankcell & 269 (13.9) & 31 (17.2) & 238 (13.5) & \blankcell \\
& Unknown
  & 1201 (10.4) & 98 (11.4) & 1103 (10.4)
  & \blankcell & 247 (12.8) & 13 (7.2) & 234 (13.4) & \blankcell \\
%
% -----------------------------
% Alcohol use
% -----------------------------
\SetCell[c=2]{l}Alcohol use && & & & $<$0.001 & & & & $<$0.001 \\
& Never / former drinker
& 10701 (93.0) & 768 (89.5) & 9933 (93.3)
& \blankcell & 1664 (86.1) & 135 (75.0) & 1529 (87.3) & \blankcell \\
& Current drinker
& 424 (3.7) & 72 (8.4) & 352 (3.3)
& \blankcell & 268 (13.9) & 45 (25.0) & 223 (12.7) & \blankcell \\
& Unknown
& 385 (3.3) & 18 (2.1) & 367 (3.4)
& \blankcell & 0 (0.0) & 0 (0.0) & 0 (0.0) & \blankcell \\
\SetCell[c=2]{l}Previous fracture &&
960 (8.3) & 304 (35.4) & 656 (6.2) & $<$0.001 &
218 (11.3) & 58 (32.2) & 160 (9.1) & $<$0.001 \\
%
% -----------------------------
% DXA Result
% -----------------------------
\SetCell[c=2]{l}DXA Result &&
& & & &
& & & \\
& Lumbar spine (L1--L4) T-score, mean (SD)
  & -1.09 (1.02) & -1.23 (1.09) & -1.08 (1.01) & $<$0.001
  & -0.74 (1.35) & -0.80 (1.22) & -0.73 (1.36) & 0.472 \\
& Femoral neck T-score, mean (SD)
  & -1.61 (0.62) & -1.80 (0.68) & -1.60 (0.62) & $<$0.001
  & -1.49 (0.73) & -1.55 (0.60) & -1.49 (0.74) & 0.195 \\
& Total hip T-score, mean (SD)
  & -1.22 (0.76) & -1.51 (0.82) & -1.20 (0.76) & $<$0.001
  & -0.87 (0.83) & -0.97 (0.85) & -0.86 (0.82) & 0.111 \\
%
% -----------------------------
% Comorbidities
% -----------------------------
\SetCell[c=2]{l}Comorbidities &&
& & & &
& & & \\
& Alzheimer's disease & 76 (0.7) & 16 (1.9) & 60 (0.6) & $<$0.001 & 9 (0.5) & 0 (0.0) & 9 (0.5) & 0.697 \\
& Ankylosing spondylitis & 18 (0.2) & 3 (0.3) & 15 (0.1) & 0.298 & 5 (0.3) & 1 (0.6) & 4 (0.2) & 0.958 \\
& Asthma & 1389 (12.1) & 126 (14.7) & 1263 (11.9) & 0.017 & 309 (16.0) & 42 (23.3) & 267 (15.2) & 0.007 \\
& Celiac disease & 131 (1.1) & 13 (1.5) & 118 (1.1) & 0.360 & 13 (0.7) & 2 (1.1) & 11 (0.6) & 0.782 \\
& Chronic kidney disease & 654 (5.7) & 85 (9.9) & 569 (5.3) & $<$0.001 & 271 (14.0) & 25 (13.9) & 246 (14.0) & 1.000 \\
& Chronic obstructive pulmonary disease & 441 (3.8) & 68 (7.9) & 373 (3.5) & $<$0.001 & 211 (10.9) & 26 (14.4) & 185 (10.6) & 0.143 \\
& Cirrhosis & 244 (2.1) & 23 (2.7) & 221 (2.1) & 0.288 & 48 (2.5) & 7 (3.9) & 41 (2.3) & 0.308 \\
& Crohn's disease & 153 (1.3) & 12 (1.4) & 141 (1.3) & 0.977 & 27 (1.4) & 5 (2.8) & 22 (1.3) & 0.186 \\
& Cushing syndrome & 21 (0.2) & 4 (0.5) & 17 (0.2) & 0.108 & 4 (0.2) & 0 (0.0) & 4 (0.2) & 1.000 \\
& Cystic fibrosis & 14 (0.1) & 1 (0.1) & 13 (0.1) & 1.000 & 1 (0.1) & 0 (0.0) & 1 (0.1) & 1.000 \\
& Eating disorder & 53 (0.5) & 7 (0.8) & 46 (0.4) & 0.181 & 4 (0.2) & 1 (0.6) & 3 (0.2) & 0.826 \\
& Hyperparathyroidism & 637 (5.5) & 35 (4.1) & 602 (5.7) & 0.063 & 77 (4.0) & 3 (1.7) & 74 (4.2) & 0.142 \\
& Hypogonadism & 157 (1.4) & 16 (1.9) & 141 (1.3) & 0.245 & 18 (0.9) & 4 (2.2) & 14 (0.8) & 0.137 \\
& Premature menopause & 772 (6.7) & 36 (4.2) & 736 (6.9) & 0.003 & 239 (12.4) & 20 (11.1) & 219 (12.5) & 0.674 \\
& Multiple myeloma & 125 (1.1) & 18 (2.1) & 107 (1.0) & 0.005 & 11 (0.6) & 3 (1.7) & 8 (0.5) & 0.125 \\
& Parkinson's disease & 85 (0.7) & 20 (2.3) & 65 (0.6) & $<$0.001 & 11 (0.6) & 1 (0.6) & 10 (0.6) & 1.000 \\
& Rheumatoid arthritis & 401 (3.5) & 46 (5.4) & 355 (3.3) & 0.003 & 114 (5.9) & 18 (10.0) & 96 (5.5) & 0.022 \\
& Stroke & 462 (4.0) & 70 (8.2) & 392 (3.7) & $<$0.001 & 66 (3.4) & 10 (5.6) & 56 (3.2) & 0.149 \\
& Systemic lupus erythematosus & 109 (0.9) & 8 (0.9) & 101 (0.9) & 1.000 & 37 (1.9) & 4 (2.2) & 33 (1.9) & 0.976 \\
& Type 1 diabetes & 122 (1.1) & 11 (1.3) & 111 (1.0) & 0.626 & 49 (2.5) & 5 (2.8) & 44 (2.5) & 1.000 \\
& Type 2 diabetes & 1736 (15.1) & 172 (20.0) & 1564 (14.7) & $<$0.001 & 520 (26.9) & 62 (34.4) & 458 (26.1) & 0.021 \\
& Ulcerative colitis & 184 (1.6) & 20 (2.3) & 164 (1.5) & 0.102 & 26 (1.3) & 1 (0.6) & 25 (1.4) & 0.531 \\
%
% -----------------------------
% Medications
% -----------------------------
% \SetCell[c=2]{l}Medications &&
% & & & &
% & & & \\
%
\SetCell[c=2]{l}\textit{Osteoporosis treatment} &&
& & & &
& & & \\
& Abaloparatide & 139 (1.2) & 27 (3.1) & 112 (1.1) & $<$0.001 & 0 (0.0) & 0 (0.0) & 0 (0.0) & \blankcell \\
& Bisphosphonates & 1759 (15.3) & 179 (20.9) & 1580 (14.8) & $<$0.001 & 294 (15.2) & 33 (18.3) & 261 (14.9) & 0.266 \\
& Denosumab & 506 (4.4) & 41 (4.8) & 465 (4.4) & 0.630 & 85 (4.4) & 8 (4.4) & 77 (4.4) & 1.000 \\
& Romosozumab & 33 (0.3) & 3 (0.3) & 30 (0.3) & 0.979 & 4 (0.2) & 0 (0.0) & 4 (0.2) & 1.000 \\
& Teriparatide & 139 (1.2) & 27 (3.1) & 112 (1.1) & $<$0.001 & 13 (0.7) & 2 (1.1) & 11 (0.6) & 0.782 \\
\SetCell[c=2]{l}\textit{Medication with negative skeletal potential} &&
& & & &
& & & \\
& Antiretrovirals & 772 (6.7) & 61 (7.1) & 711 (6.7) & 0.675 & 28 (1.4) & 1 (0.6) & 27 (1.5) & 0.468 \\
& Aromatase inhibitors & 1034 (9.0) & 55 (6.4) & 979 (9.2) & 0.007 & 121 (6.3) & 11 (6.1) & 110 (6.3) & 1.000 \\
& Calcineurin inhibitors & 1156 (10.0) & 104 (12.1) & 1052 (9.9) & 0.041 & 85 (4.4) & 8 (4.4) & 77 (4.4) & 1.000 \\
& Enzyme-inducing antiepileptic drugs & 141 (1.2) & 19 (2.2) & 122 (1.1) & 0.010 & 27 (1.4) & 5 (2.8) & 22 (1.3) & 0.186 \\
& {\small GnRH antagonist} & 281 (2.4) & 30 (3.5) & 251 (2.4) & 0.049 & 4 (0.2) & 1 (0.6) & 3 (0.2) & 0.826 \\
& Systemic steroids & 2649 (23.0) & 261 (30.4) & 2388 (22.4) & $<$0.001 & 1277 (66.1) & 126 (70.0) & 1151 (65.7) & 0.281 \\
\end{longtblr}

%% file: tables/tab_fracture_cat.tex
\begin{tabularx}{.8\textwidth}{lXrrrr}
\toprule
\textbf{Fracture Category} &
\textbf{ICD-10 Code} &
\multicolumn{2}{r}{\textbf{NYP/WCM}} &
\multicolumn{2}{r}{\textbf{INPC}}\\
\cmidrule(l){3-4}\cmidrule(l){5-6}
& & \textbf{No} & \textbf{\%} & \textbf{No} & \textbf{\%}\\
\midrule
Forearm, Wrist &
S52.5*, S52.6* &
111 & 1.0 & 74 & 3.8\\
\arrayrulecolor{stone}\hline
Hip, Femur &
S72.0*, S72.1*, S72.2*, S72.3*, S72.4*, S72.8*, S72.9* &
98 & 0.9 & 33 & 1.7\\
\arrayrulecolor{stone}\hline
Lower Leg, Ankle, Foot &
S82* &
150 & 1.3 & 94 & 4.9\\
\arrayrulecolor{stone}\hline
Shoulder, Humerus &
S42.0*, S42.1*, S42.2*, S42.3*, S42.4* &
124 & 1.1 & 48 & 2.5\\
\arrayrulecolor{stone}\hline
Spine, Vertebral &
M48.4*, M48.5*, M49.5*, 
S22.0*, S22.1*,
S32.0*, S32.1*, S32.2*, S32.5*, S32.7*, S32.8*,
T08*&
376 & 3.3 & 90 & 4.7\\
\arrayrulecolor{stone}\hline
No Fracture & -- & 10,663 & 92.5 & 1592 & 82.4\\
\arrayrulecolor{black}
\bottomrule
\end{tabularx}

%% file: tables/tab_variables.tex
\begin{tabular}{lcc}
\hline
\textbf{Covariates} & \textbf{\modela} & \textbf{\modelb} \\
\hline
Age, Sex \& BMI & \cmark & \cmark \\
\arrayrulecolor{stone}\hline
Smoking status \& Alcohol use & \cmark & \cmark \\\hline
$T_{\min}$ (minimum of 3 DXA T-scores) & \cmark & \xmark \\\hline
Femoral neck (FN) T-score  & \xmark & \cmark \\\hline
Comorbidities                 
& \cmark & Rheumatoid Arthritis (RA) only \\\hline
Medications with negative skeletal potential  & \cmark & Systemic Steroids only \\\hline
Osteoporosis treatments
& \cmark & \xmark \\
\arrayrulecolor{black}\hline
\end{tabular}

%% file: tables/tab_predicator.tex
\begin{tabularx}{\textwidth}{
>{\raggedright\arraybackslash}X c
>{\raggedright\arraybackslash}X c
>{\raggedright\arraybackslash}X c
>{\raggedright\arraybackslash}X c
}
\toprule
\multicolumn{2}{c}{\textbf{Cox}}
& \multicolumn{2}{c}{\textbf{XGBoost}}
& \multicolumn{2}{c}{\textbf{RSF}}
& \multicolumn{2}{c}{\textbf{GBS}} \\
\cmidrule(lr){1-2}\cmidrule(lr){3-4}\cmidrule(lr){5-6}\cmidrule(lr){7-8}
\textbf{Feature} & \textbf{Mean}
& \textbf{Feature} & \textbf{Mean}
& \textbf{Feature} & \textbf{Mean}
& \textbf{Feature} & \textbf{Mean} \\
\midrule

\rowcolor{gray!15}\multicolumn{8}{l}{\textit{\modela}} \\
Previous fracture & 0.1265 & Previous fracture & 0.1435 & Previous fracture & 0.1304 & Previous fracture & 0.1317 \\
Age              & 0.0445 & Age              & 0.0166 & Age              & 0.0414 & Age              & 0.0448 \\
Minimum T-score  & 0.0260 & Minimum T-score  & 0.0133 & Minimum T-score  & 0.0131 & Minimum T-score  & 0.0140 \\
Alcohol status   & 0.0075 & Comorbidities    & 0.0066 & Comorbidities    & 0.0076 & Alcohol status   & 0.0076 \\
Comorbidities    & 0.0065 & Sex              & 0.0035 & Bone-protective meds & 0.0040 & Comorbidities & 0.0071 \\
\midrule

\rowcolor{gray!15}\multicolumn{8}{l}{\textit{\modelb}} \\
Previous fracture    & 0.1294 & Previous fracture    & 0.1497 & Previous fracture    & 0.1391 & Previous fracture    & 0.1343 \\
Age                  & 0.0470 & Age                  & 0.0126 & Age                  & 0.0424 & Age                  & 0.0499 \\
Femoral neck T-score & 0.0146 & Femoral neck T-score & 0.0048 & Femoral neck T-score & 0.0098 & Steroid count        & 0.0088 \\
Steroid count        & 0.0123 & Steroid count        & 0.0045 & Body mass index      & 0.0058 & Femoral neck T-score & 0.0081 \\
Alcohol status       & 0.0080 & Body mass index      & 0.0025 & Steroid count        & 0.0023 & Alcohol status       & 0.0081 \\
\bottomrule
\end{tabularx}

%% file: suppl.tex
\section{Rule-based extraction of DXA results from radiology reports}
\label{sec:nlp}

DXA measurements were extracted from free-text radiology reports using a deterministic rule-based natural language processing pipeline implemented in Python with regular expressions. The goal was to convert narrative DXA reports into structured variables suitable for modeling.

\subsection{Identification of report sections and DXA measurements}

Each DXA report was parsed line by line to identify the major anatomic sections, including the lumbar spine, left and right femurs, and left and right forearms. Once a section was identified, subsequent measurement lines were interpreted within that anatomic context.

Measurement rows were extracted using regular-expression patterns designed to capture region name, BMD, T-score, and Z-score. Numeric values were cleaned and converted to floating-point format. Region labels were standardized to reduce lexical variation. For ambiguous labels such as ``neck'' or ``total,'' the parent anatomic section was appended to preserve specificity (e.g., ``NECK (LEFT FEMUR)'').

\subsection{Extraction of FRAX values}

When present in the report, FRAX probabilities were extracted using keyword-based rules targeting major osteoporotic fracture risk and hip fracture risk fields, including common expressions such as ``FRAX,'' ``major osteoporotic'' and ``hip.''

\subsection{Structured output and quality control}

Each successfully parsed region-level measurement generated 1 structured record containing note-level identifiers and extracted numeric values. Reports or rows that could not be reliably parsed were excluded. T-scores outside a clinically plausible range were manually reviewed against source reports and corrected when possible; unresolved errors were excluded.

This approach was efficient and interpretable but dependent on the semi-structured formatting of DXA reports in the contributing health systems. Reports using substantially different layouts or phrasing may require additional pattern refinement.

% \subsection{Quality control and limitations}

% The extraction approach was deterministic and designed for the semi-structured format of DXA radiology reports in our health system. Rows that could not be parsed into valid numeric values were excluded. T-scores outside a clinically reasonable range were manually reviewed against the original DXA note and corrected when possible; unresolved errors were excluded.

% Although this regular-expression approach was efficient and interpretable, its performance depended on the consistency of note formatting and terminology. Reports with substantially different layouts or uncommon phrasing may require additional pattern refinement.

\newpage

\section{Permutation importance}
\label{sec:permutation}

To examine predictor contributions across model classes, we assessed permutation importance on the evaluation data. Feature importance was defined as the reduction in model discrimination after permuting the values of a single feature~\cite{fisher2019all}. Importance estimates were obtained within each internal-validation split and then aggregated across splits.

Permutation importance analyses showed a consistent pattern across model classes (\textbf{eTable~\ref{tab:perm_top5_internal_4models}}). Prior fracture history was the highest-ranked predictor, followed by age. The minimum T-score consistently emerged as a leading predictor across Cox, XGBoost, RSF, and GBS. Comorbidity burden also ranked highly, and alcohol use and osteoporosis treatment history emerged as important features in selected models. 